\documentclass[
    ,final            
  ]
  {aipproc}

\layoutstyle{6x9}

\begin{document}

\title{Swift X-ray Afterglows and the Missing Jet Break Problem}


\author{J. L. Racusin}{
  address={Department of Astronomy \& Astrophysics, The Pennsylvania State
  University, 525 Davey Lab, University Park, PA 16802}
}

\author{E.-W. Liang}{
  address={Department of Physics, University of Nevada, Las Vegas, NV 89154},
  altaddress={Department of Physics, Guangxi University, Nanning 530004, China}
}

\author{D. N. Burrows}{
  address={Department of Astronomy \& Astrophysics, The Pennsylvania State
  University, 525 Davey Lab, University Park, PA 16802}
}

\author{A. Falcone}{
  address={Department of Astronomy \& Astrophysics, The Pennsylvania State
  University, 525 Davey Lab, University Park, PA 16802}
}

\author{D. C. Morris}{
  address={Department of Astronomy \& Astrophysics, The Pennsylvania State
  University, 525 Davey Lab, University Park, PA 16802}
}

\author{B. B. Zhang}{
  address={Department of Physics, University of Nevada, Las Vegas, NV 89154}
}

\author{B. Zhang}{
  address={Department of Physics, University of Nevada, Las Vegas, NV 89154}
}

\begin{abstract}
We present a systematic survey of the temporal and spectral properties of
all GRB X-ray afterglows observed by Swift-XRT between January 2005 and July
2007.  We have constructed a catalog of all light curves and spectra and
investigate the physical origin of each afterglow segment in the
framework of the forward shock models by comparing the data with the
closure relations.  We search for possible jet-like breaks in the
lightcurves and try to explain some of the "missing" X-ray jet breaks in
the lightcurves.
\end{abstract}

\maketitle

\section{Introduction}

Studies of the presence or absence of jet breaks in GRB X-ray afterglows
have  
been recently undertaken with several different approaches yielding differing
results \citep{burrows07,liang07b,panaitescu07,kocevski07}.  The importance of
this work lies in that the results have
vital implications on the energetics, geometry, and frequency of GRBs.  The fact
that they do not behave as expected from pre-{\it Swift} observations is not
surprising in the context of how much we do not understand and have only
recently learned about all of the aspects of X-ray afterglows.  To understand
the jet break phenomena we must understand it in the global context of GRB and
afterglow properties.  Therefore, the goal of our study is to do a census of
X-ray afterglow properties by fitting a variety of physical models to each
component of the afterglows and understand how the jet breaks fit in as one
component in the larger coherent picture of this phenomenon.

\section{Analysis}
Our sample consists of all GRBs observed by {\it Swift}-XRT between January
2005 and July 2007 with enough counts to make and fit light curves and spectra. Our
resulting sample contains 212 X-ray afterglows, 14 of which were not originally
discovered by Swift-BAT, and 80 of which have redshifts.  We created light curves
for each afterglow using the Penn State XRT light curve tools, removing all
significant flares, and fit them to power-laws and (multiply-) broken power-laws.

We attempt to categorize these light curves in the context of the canonical
model \citep{zhang06,nousek06}.  The canonical model contains 5 segments; I:
the initial steep decay often referred to as the high-latitude emission or 
curvature effect \citep{zhang07a}
; II: the plateau 
which is believed to be due to continuous energy injection from the central
engine \citep{liang07a}
; III: the normal decay due to adiabatic evolution of the forward shock
\citep{meszaros02}
; IV: the post-jet break phase
\citep{rhoads99,sari99}
; V: flares, which are 
seen in $\sim 1/3$ of all {\it Swift} GRB X-ray afterglows and are believed to be
caused by continuous sporadic emission from the central engine
\citep{chincarini07,falcone07}.  We classify the light curves depending 
upon criteria of the number of segments and their relative decay indices,
leading to unambiguous categories of segments I-II-III-IV and II-III-IV that
contain jet breaks in IV, segments I-II and I-II-III that are apparently
pre-jet break with some ambiguity in the segments III, the ambiguous segments
II-III/III-IV, and single power-laws. The ambiguous groups may well contain many
of the missing jet breaks and require further distinguishing criteria.

To further investigate the properties of the straightforward jet breaks and the
ambiguous cases, we created spectra for each of these segments of the light
curves and fit them to absorbed power-laws.  These temporal and spectral
properties are used in conjunction to characterize the afterglows.

The closure relations describe the temporal and spectral evolution of the
afterglows with dependence on the physical mechanisms at work in the GRB
and its environment.  We assembled many permutations of the closure relations
that depend on the circum-GRB environment, the frequency regime, slow or fast
cooling, electron spectral index regime, presence of energy injection,
isotropic or collimated emission, and 
jet structure from the literature
\citep{zhang04,zhang06,dai01,panaitescu06a,panaitescu05b}.  We applied these
relations to each light curve 
segment, where appropriate, using the compiled temporal and spectral indices.  The
resulting fits allowed us to distinguish those light curves with potential jet
breaks that are consistent with the post-jet break closure relations and those
that are not.  We require closure relation consistency between the segments of
each light curve and use the corresponding information to eliminate models
that cannot appropriately be applied throughout.  Unfortunately, due to the large
number of possible models, often many relations were consistent and further
distinguishing criteria were required.

\section{Results}
\begin{figure}
  \includegraphics[height=.35\textheight,angle=90]{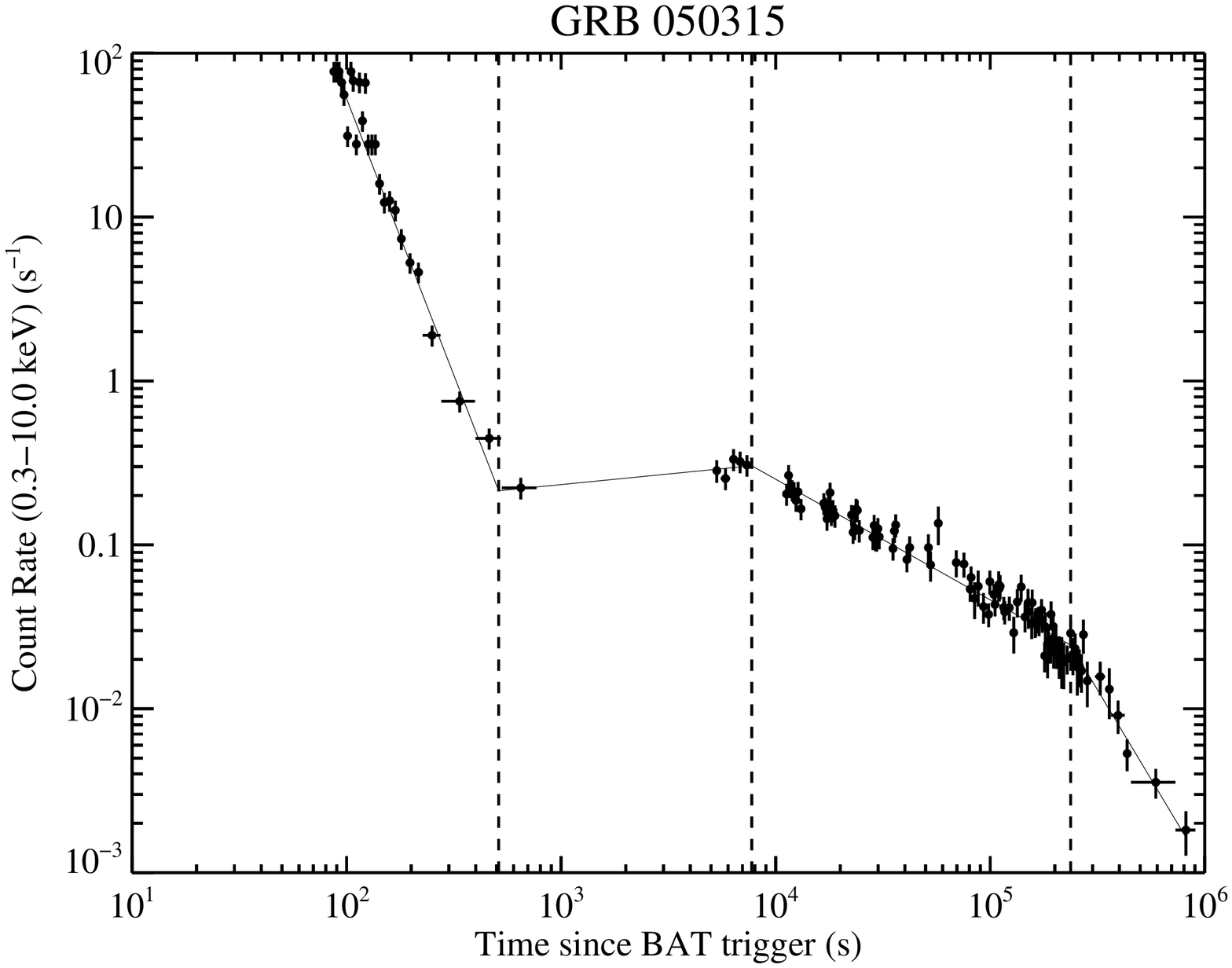}
  \includegraphics[height=.35\textheight,angle=90]{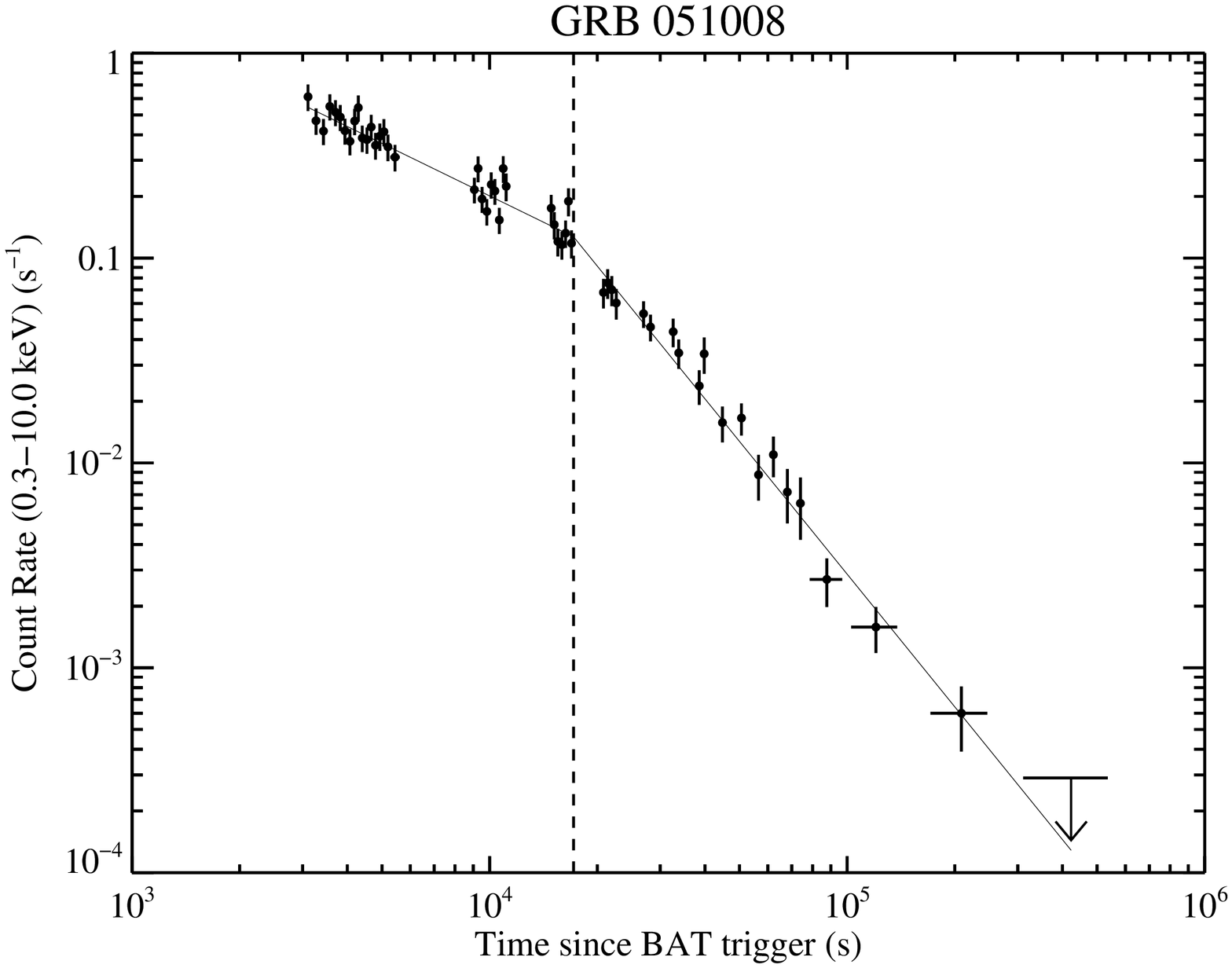}
  \caption{{\it Left} - Example of Prominent jet break in GRB 050315 with fit
    showing all 4 segments. {\it Right} - Example of ambiguous 2 segment light
    curve for GRB 051008 with a Probable jet
    break classified using $\alpha$ comparison technique. \label{fig:examples}}
\end{figure}

Using the temporal fit criteria and the closure relations fits, we classified
our sample into several categories of potential jet breaks based upon their
likelihood of containing jet breaks.  Those afterglow light curves that
distinctly contain a segment IV that is consistent with at least one post-jet
break closure relations are categorized as Prominent jet breaks and constitute
$\sim 13\%$ of our total sample.  GRB 050315 (shown in the left panel of Figure
\ref{fig:examples}) is an example of a burst in this category.  Those
ambiguous light curves (segments I-II-III, II-III, single power-laws) that are
consistent with only post-jet break and not pre-jet break closure relations are
categorized as Hidden jet breaks, which constitute $\sim 3\%$ or our total sample.
The remaining temporally ambiguous light 
curves that are consistent with both pre- and post-jet break closure relations
require further distinction.

To further distinguish post-jet break from pre-jet break light curves we compare
the relative decay slopes of the apparent II-III transition of the ambiguous
sample to the Prominent jet break sample.  Though this technique we find that an
additional $25\%$ of our sample contain apparent jet break transitions like that
of GRB 051008 shown in the right panel of Figure \ref{fig:examples}.

For those ambiguous light curves that do not contain even the apparent II-III
transition, namely the single power-laws, we 
evaluate their temporal decay slopes and start and stop times relative to the
Prominent jet breaks, finding that most of those with steep decays begin during
the time frame where we would expect to find jet breaks.  Therefore, those that
start late and are steep are probably post-jet break and those that start early
and end early are probably pre-jet break.  Through these criteria we suggest
that an
additional $\sim 3\%$ of our sample contain jet breaks or are post-jet break.

\begin{figure}
  \includegraphics[height=.35\textheight,angle=90]{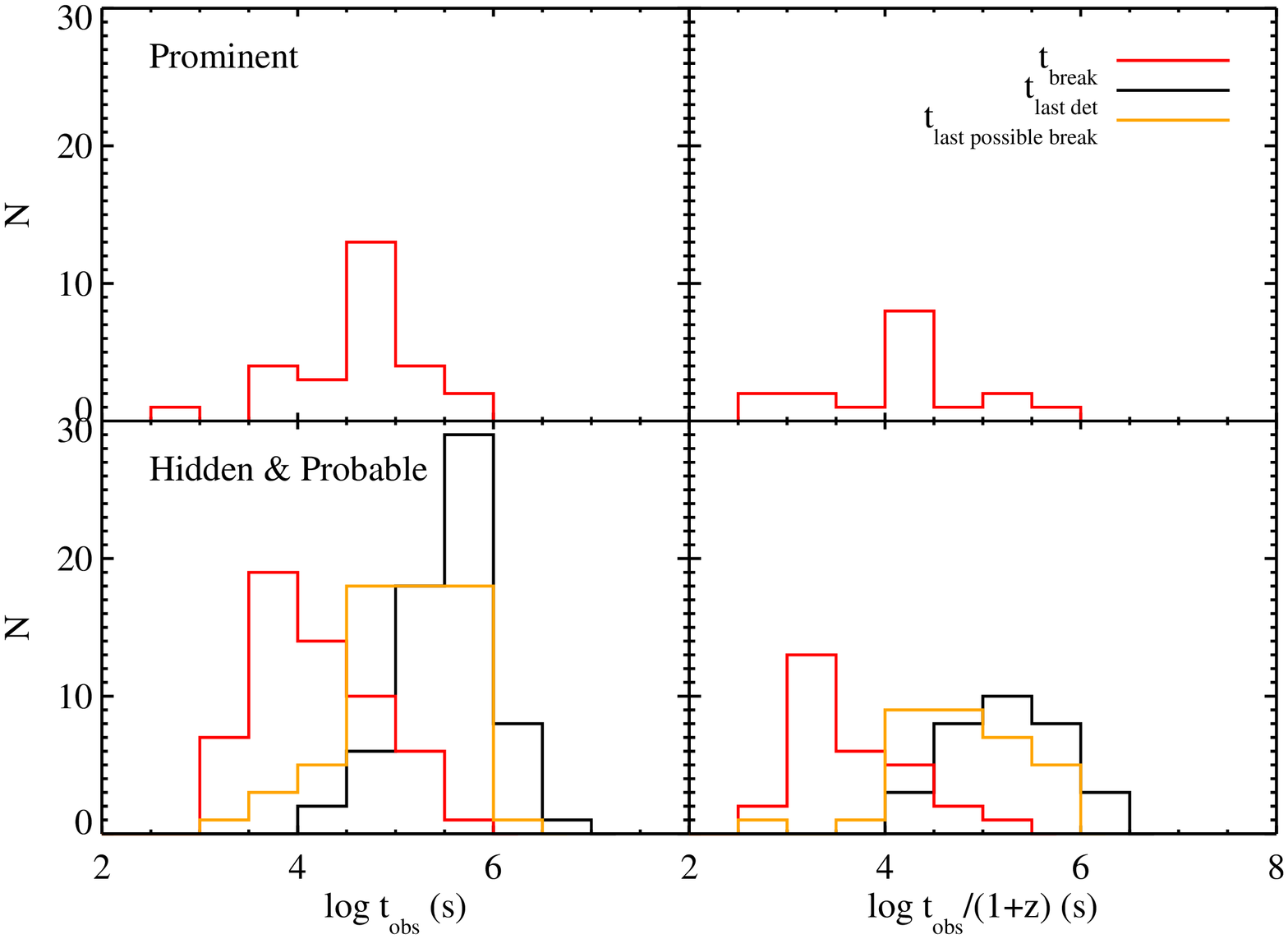}
  \includegraphics[height=.35\textheight,angle=90]{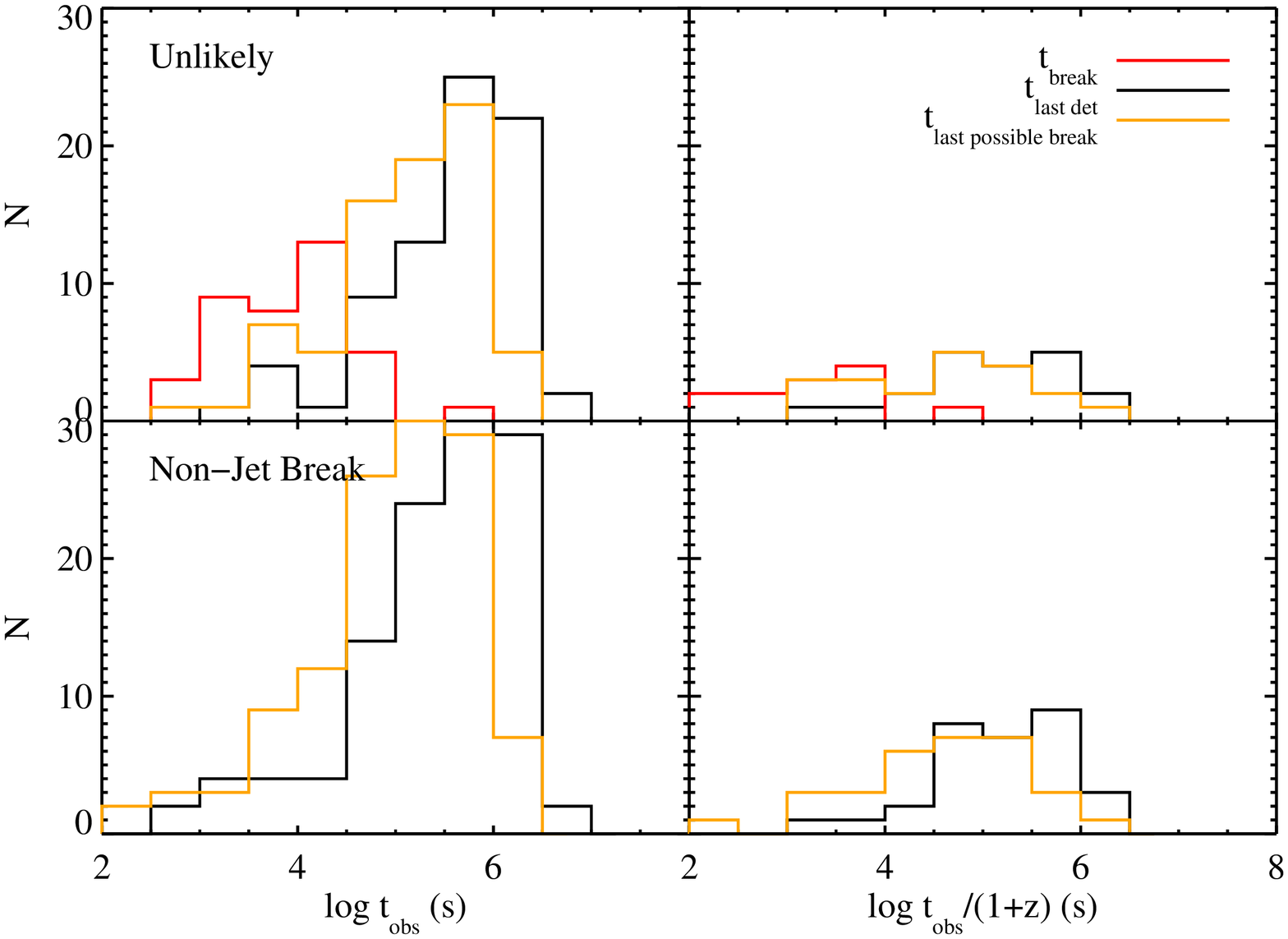}
  \caption{Distributions of potential jet break times, time of last detection,
  and time of last possible break for our categories of light curves with
  potential jet breaks and non-jet breaks in the observed and rest frame for
  those with redshifts. 
  \label{fig:tlastdet}}
\end{figure}

The other logical explanation for not seeing jet breaks for every GRB is that
the observations simply end too early.  We evaluate this by calculating the
last time for which a break could occur and be buried within the errors.  If
this time is inside the time interval for which we expect a jet break to occur
based upon the behavior the Prominent sample, then it is feasible that a jet
break occurred around or after this time and would still be consistent with
expectations.  We compare these distributions in Figure \ref{fig:tlastdet}, and
find that these criteria are met for $\sim 80\%$ of the remaining afterglows.

\section{Conclusions}

Although the jet break phenomena is not fully understood, we are beginning to
be able to better explain the paucity of expected observations of them.  We are
able to identify a sizeable group of afterglows that likely contain jet breaks or
post-jet break data even though they do not present themselves in the classical
context of the full canonical model.  Due to observational limitations and
additional possible physical model variations, even more afterglows may be
consistent with the expectations from those that do contain confident jet
breaks, but are currently indistinguishable. While we are beginning to
understand or at least be able to explain the majority of our sample, we are
also finding an interesting small subset of outliers that confidently do not
contain jet breaks during the time interval in which we would expect to see
them.  These afterglows require further investigation and perhaps are somehow
fundamentally different in their jet and afterglow properties.

\begin{theacknowledgments}
JLR, DNB, DCM, and AF acknowledge support under NASA contract NAS5-00136.
\end{theacknowledgments}

\bibliographystyle{aipproc}   

\bibliography{ms}

\end{document}